\pdfoutput=1
\documentclass[9pt,twocolumn,twoside]{opticajnl}
\journal{opticajournal}
\doi{}
\setboolean{shortarticle}{true}
\setboolean{displaycopyright}{false}
\usepackage{amsmath,amssymb,bm,graphicx,xcolor}

\newcommand{\kw}{k_\mathrm{w}}

\title{Levitated near-field optical traps above waveguides}
\author[1,2,*]{Jinsheng Lu}
\affil[1]{School of Power and Mechanical Engineering, Wuhan University, Wuhan 430072, China}
\affil[2]{School of Integrated Circuits, Wuhan University, Wuhan 430072, China}
\affil[*]{jinslu@whu.edu.cn}
\begin{abstract}
Evanescent fields of guided modes decay away from a waveguide, so near-field optical traps traditionally hold particles against its surface. Here, we show that an intensity maximum can be lifted off the waveguide surface by superposing two guided modes of different order, each forming a standing wave. We derive a simple criterion for this lifted maximum: the product of the two mode decay constants must be smaller than the squared wavenumber of the surrounding medium. Theoretical modeling, supported by full-wave simulations, predicts that a silicon nitride waveguide suspended in water holds a silicon sphere of radius 100~nm at a height of 0.4~$\mu$m above its surface, with a trapping potential depth of 0.18~$k_\mathrm{B}T$ per milliwatt of total guided power. The power ratio of the two modes tunes this height. Such traps could enable contact-free studies of nanoscale dynamics and interactions near surfaces.
\end{abstract}
\begin{document}
\maketitle

Optical forces can accelerate, levitate, pull and trap small particles \cite{Ashkin1986,Lu2024LSA}, and optical tweezers have become a standard tool in physics and biology, including on chips for high-throughput manipulation \cite{He2026}. Near-field traps confine the light below the diffraction limit and hold smaller objects \cite{Juan2011,Marago2013,Zhang2021}. Evanescent fields propel particles along a surface \cite{Kawata1992}, and on photonic chips the evanescent fields of waveguides and resonators hold and transport them \cite{Kawata1996,Yang2009,Lin2010,Soltani2014}, with forces and torques that also depend on the helicity and polarization of the field \cite{Lu2023helicity,Lu2023microfiber}, and guided modes induce optical binding between particles \cite{Toftul2020}. Beams that metalenses or optical phased arrays radiate from a chip trap particles far above its surface \cite{Shen2021,Yu2024,Sneh2024}, but near-field traps stay at the surface. Because the evanescent field decays away from the surface, the gradient force pulls a particle of positive polarizability toward the waveguide \cite{Ng2000}, and the trap sites lie on the surface, as in the standing-wave lattices of counter-propagating evanescent waves \cite{Siler2006,Soltani2014}. Contact brings friction, adhesion and a hydrodynamic drag that depends on the gap \cite{Brenner1961,Schaffer2007}. A trap that holds the particle at a height above the surface would remove these complications.

\begin{figure}[t]
\centering\includegraphics[width=\columnwidth]{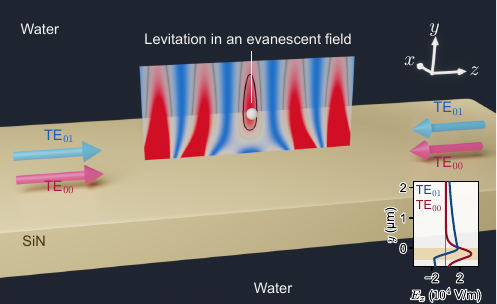}
\caption[Levitated near-field trap above a waveguide.]{Levitated near-field trap above a waveguide. The TE$_{00}$ and TE$_{01}$ modes of a wide silicon nitride waveguide suspended in water are launched from both ends, and their evanescent tails interfere above the surface. Cut plane: computed $|E|^2$, with the color scale saturated at the lifted maximum, which lies about 0.4~$\mu$m above the surface; the particle is drawn small, not to scale. Inset: field $E_x$ of TE$_{00}$ (red) and TE$_{01}$ (blue) against the height $y$ across a 370-nm-thick waveguide at 1064 nm wavelength, for 1 W per meter of width (slab modes).}
\label{fig:schematic}
\end{figure}

Interference of guided modes can change the decay of the near field. Modes of different order have different decay constants above the surface, so their superposition can grow with height before it decays, and the intensity then peaks above the surface, at what we call a lifted maximum. Such intensity landscapes have been designed with cascaded mode converters, but the lifted maximum extends over only a short stretch of waveguide and reaches only a small fraction of the surface intensity \cite{Ginis2020}. In holographic superpositions of evanescent waves, the intensity peaks along the height only at saddle points \cite{Thomson2008}. Atoms have been held above surfaces in evanescent light of two colors, one red- and one blue-detuned, so that the two dipole potentials have opposite signs \cite{Ovchinnikov1991,Vetsch2010,Balykin2014}, and Bloch surface waves at two frequencies can hold a Mie-resonant silicon particle near a surface \cite{Shilkin2022}. At a single frequency, interfering guided modes trap atoms of negative polarizability at intensity minima \cite{Christandl2004,Sague2008}. To our knowledge, no particle of positive polarizability has been levitated by a single-frequency evanescent field, and the conditions under which superposed guided modes can hold such a particle above a waveguide are not known.

Here, we show that a wavenumber budget set by the Helmholtz equation decides when superposed evanescent waves can hold a particle above a waveguide. The modes must form standing waves, because co-propagating modes can raise the intensity along the height only at a saddle point, and for two such modes the budget gives a simple criterion: a trap forms above the surface exactly when the product of the two decay constants is smaller than the squared wavenumber of the surrounding medium. On silica, the particle escapes along the waveguide over a shallow saddle; a silicon nitride waveguide suspended in water instead holds a silicon nanosphere in a trap 20 times deeper.

In Fig.~\ref{fig:schematic}, the two modes of a thin waveguide, TE$_{00}$ and TE$_{01}$ (TE$_{mn}$ has $m$ field nodes across the width and $n$ across the thickness), are each launched from both ends so that they form standing waves, and their interference lifts an intensity maximum off the surface. A maximum along the height is not yet a trap: the intensity must also fall off along and across the waveguide, and the interference that lifts the maximum in one direction tends to lower it in another. We consider the surrounding medium above a waveguide, homogeneous with refractive index $n_\mathrm{w}$ (water in our examples), where every Cartesian component $E_j$ of the monochromatic electric field obeys the Helmholtz equation $\nabla^2E_j+\kw^2E_j=0$, with $\kw=n_\mathrm{w}k_0$. At a local maximum of $|E_j|$ of a real field component (we take $E_j>0$ there) the field is flat and curves downward in every direction, so the local wavenumbers defined by $k_x^2=-\partial_x^2E_j/E_j$, and likewise for $y$ and $z$, are real. The Helmholtz equation then fixes the sum of their squares,
\begin{equation}
k_x^2+k_y^2+k_z^2=\kw^2,\qquad k_x^2,\,k_y^2,\,k_z^2>0 .
\label{eq:budget}
\end{equation}
This is the dispersion relation of a plane wave, now imposed on the local wavenumbers at the maximum, which measure the curvature of the field there rather than the components of a wavevector, and we call it the wavenumber budget: the squared local wavenumbers measure how sharply the field falls off from the maximum in each direction, their sum is fixed, and a sharper maximum in one direction is a broader one in the others.

\begin{figure}[t]
\centering\includegraphics{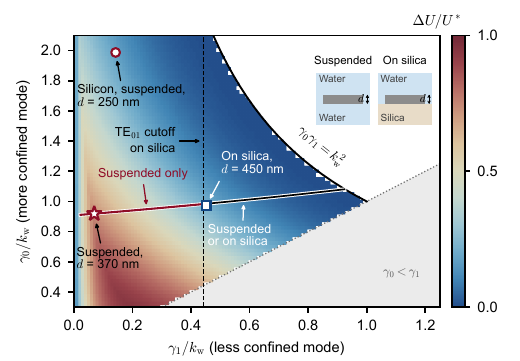}
\caption[Levitation criterion and relative trapping potential depth against the two decay constants.]{Levitation criterion and relative trapping potential depth against the decay constants $\gamma_0$ and $\gamma_1$ of the more and the less confined mode, in units of $k_\mathrm{w}$; waveguides in water, silicon nitride at 1064~nm and silicon at 1550~nm. Solid line: the criterion $\gamma_0\gamma_1=k_\mathrm{w}^2$, Eq.~(\ref{eq:criterion}); a lifted maximum exists only below it. Color: relative trapping potential depth $\Delta U/U^\ast$ of a point dipole. Dashed line: cutoff of TE$_{01}$ on a silica substrate, $\gamma_1/k_\mathrm{w}=0.44$. White-edged line: suspended silicon nitride waveguides, thicker to the right, from the cutoff of TE$_{01}$ ($d=355$~nm) to $d=1.0~\mu$m, where the line meets the criterion; waveguides on silica lie on its black part, from $d=436$~nm. Square: waveguide on silica, $d=450$~nm; star: suspended waveguide, $d=370$~nm; circle: suspended silicon waveguide, $d=250$~nm. Inset: the two cross sections, with the thickness $d$.}
\label{fig:map}
\end{figure}

Above the surface the field of guided modes is a sum of evanescent waves \cite{NovotnyHecht}. Let $x$ be the coordinate across the width, $y$ the height and $z$ the direction of propagation. When each mode is present as an equal-amplitude counter-propagating pair and the standing waves oscillate in phase, the field at a fixed lateral position is
\begin{equation}
E_x(y,z)=\sum_i A_i\,e^{-\gamma_i y}\cos(\beta_i z+\varphi_i),\qquad \gamma_i=\sqrt{\beta_i^2-\kw^2},
\label{eq:field}
\end{equation}
where $\beta_i$ is the propagation constant and $\gamma_i$ the decay constant of mode $i$; we write the component as $E_x$, the only one for the TE modes considered below. $A_i$ includes the lateral profile of the mode; this factorization is exact for slab modes and holds for a waveguide much wider than $1/\kw$. Each term has $k_y^2=-\gamma_i^2$ and $k_z^2=\beta_i^2$, so the field obeys Eq.~(\ref{eq:budget}) with $k_x=0$, and a maximum, with $k_y^2>0$ and $k_z^2>0$, needs
\begin{equation}
0<k_y^2<\kw^2 .
\label{eq:necessary}
\end{equation}
For a single evanescent wave $k_y^2=-\gamma^2<0$, an imaginary local wavenumber; at a lifted maximum, where evanescent waves of opposite sign interfere, $k_y^2$ must become positive and stay below $\kw^2$, for any number of modes.

For two modes, a more confined one with decay constant $\gamma_0$ and a less confined one with $\gamma_1<\gamma_0$, the vertical condition $\partial_yE_x=0$ gives $\gamma_0E_0=-\gamma_1E_1$: the two terms have opposite signs, the more confined mode cancels the less confined one below the maximum, and the less confined mode dominates at the maximum. At the antinodes of both standing waves, where the height and the propagation direction decouple, this fixes the local wavenumbers, $k_y^2=\gamma_0\gamma_1$ and, by Eq.~(\ref{eq:budget}), $k_z^2=\kw^2-\gamma_0\gamma_1$ (Supplement 1, Sec.~S1), so the field has a maximum exactly when
\begin{equation}
\gamma_0\gamma_1<\kw^2 .
\label{eq:criterion}
\end{equation}
Equation~(\ref{eq:criterion}) is the central result of this Letter: the vertical confinement of the lifted maximum takes the share $\gamma_0\gamma_1$ of the budget, and the remainder left for the direction of propagation has to be positive. Equivalently, the geometric mean of the two decay lengths exceeds $1/\kw=\lambda/(2\pi n_\mathrm{w})$, which is 128~nm in water at 1064~nm, and because $\gamma_1<\gamma_0$, the less confined mode is weakly confined, $\gamma_1<\kw$, with an effective index below $\sqrt2\,n_\mathrm{w}$. The intensity at the lifted maximum is
\begin{equation}
I^\ast=\left(1-\frac{\gamma_1}{\gamma_0}\right)^2 I_{1}(y^\ast),
\label{eq:bound}
\end{equation}
where $I_1(y^\ast)$ is the intensity of the less confined mode alone at the height $y^\ast$ of the maximum, so the two decay constants must differ for a maximum to exist. Co-propagating modes fail for any decay constants: their intensity has a single spatial frequency along $z$, and the cancellation that lifts the maximum along $y$ is the minimum of this oscillation along $z$, so the point where the field peaks along $y$ is a saddle that does not confine the particle along the waveguide (Supplement 1, Sec.~S2).

\begin{figure}[t]
\centering\includegraphics{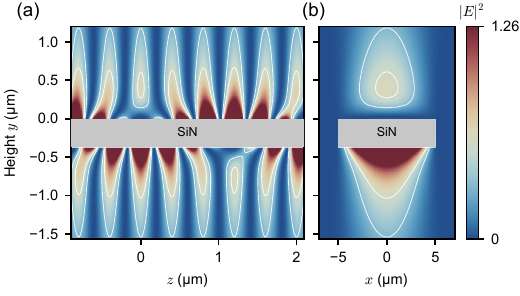}
\caption[Levitated trap above a suspended silicon nitride waveguide.]{Levitated trap above a suspended silicon nitride waveguide (gray), 370~nm thick and 10~$\mu$m wide, in water at 1064~nm, computed from the modes of the waveguide given by a finite-difference mode solver: $|E|^2$ in units of $10^{12}$~V$^2$/m$^2$ for a TE$_{00}$ to TE$_{01}$ power ratio of 0.74 and 20~mW of guided power (total of the four waves), (a) along the waveguide on its center line, $x=0$, and (b) across it through the upper trap, $z=0$, with the horizontal scale compressed eightfold. White contours: 0.5 and 0.9 of $|E|^2$ at the traps, which lie 400~nm above and below the waveguide.}
\label{fig:trap}
\end{figure}

Figure~\ref{fig:map} shows where real waveguides lie relative to the criterion. Pairs that differ in order across the thickness can levitate a particle: TE$_{00}$ and TE$_{01}$ of thin silicon nitride waveguides, on silica or suspended in water, lie below the criterion line, and so does a suspended 250-nm-thick silicon waveguide, with $\gamma_0\gamma_1/\kw^2=0.28$. Pairs that differ in order across the width fail (Supplement 1, Sec.~S4, Fig.~S1): TE$_{00}$ and TE$_{20}$ of a silicon-on-insulator waveguide 1100~nm wide and 220~nm thick in water at 1550~nm give $\gamma_0\gamma_1/\kw^2=2.1$, while TE$_{00}$ and TE$_{10}$ of a 6-$\mu$m-wide silicon nitride waveguide have nearly equal decay constants, so Eq.~(\ref{eq:bound}) leaves almost no intensity at the maximum.%

\begin{figure}[t]
\centering\includegraphics{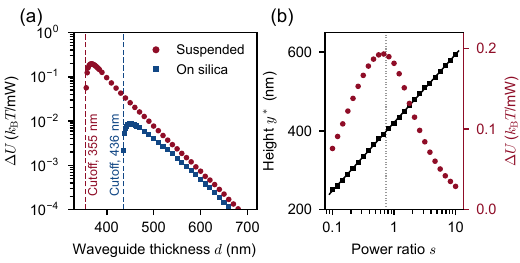}
\caption[Design of the trap: waveguide thickness and power ratio.]{Design of the trap for a silicon sphere of radius 100~nm on a 10-$\mu$m-wide silicon nitride waveguide in water at 1064~nm. (a) Trapping potential depth $\Delta U$ per milliwatt of guided power against the waveguide thickness, for a waveguide suspended in water (circles) and on silica (squares), with the power ratio chosen at each thickness for the deepest trap and the antinodes of the two in-phase standing waves aligned (slab model, effective width 5~$\mu$m); dashed lines: thickness at which TE$_{01}$ is cut off. (b) Height $y^\ast$ of the maximum (squares; line: $y^\ast(s)=y^\ast(1)+\ln s/[2(\gamma_0-\gamma_1)]$) and $\Delta U$ (circles) of the suspended 370-nm-thick waveguide against the power ratio $s$; vertical dotted line: $s=0.74$, the power ratio of Fig.~\ref{fig:trap}.}
\label{fig:design}
\end{figure}

To build a trap we take a silicon nitride waveguide ($n=2.00$) of thickness $d$ on a substrate of index $n_\mathrm{s}$ under water ($n=1.326$) at 1064~nm, with the standing TE$_{00}$ and TE$_{01}$ waves of Fig.~\ref{fig:schematic}; the guided power is the total power of the four waves. The waveguide is 10~$\mu$m wide, much wider than the wavelength, so we take its field along the height from the exact modes of an infinite slab, whose decay constants differ from those of the 10-$\mu$m-wide waveguide by less than 3\,\% (Supplement 1, Sec.~S6). In this slab model, both modes have an effective width of 5~$\mu$m, half the width of the waveguide. We use three force models (Supplement 1, Table~S3). The point dipole, with the potential $U=-\tfrac14\,\mathrm{Re}(\alpha)\,|E|^2$ \cite{Chaumet2000,NietoVesperinas2010} and $\alpha$ from the electric-dipole Mie coefficient \cite{BohrenHuffman}, gives the criterion and the relative depth; the nonconservative force, which follows the phase gradient and the spin of the field \cite{Albaladejo2009}, vanishes in this real, linearly polarized field. For a silicon sphere ($n=3.55$) of radius 100~nm we compute the exact force on the isolated sphere from Mie theory, expanding the field in vector spherical waves about the sphere \cite{Gouesbet2017}, and integrate it into a potential; all depths quoted below for the sphere use this model. The depth $\Delta U$ of the trapping potential is the barrier on the easiest escape route, that is, the lowest saddle on any path from the trap to the contact plane, where the sphere touches the waveguide, or away from the surface; we give it per milliwatt of guided power. Three-dimensional finite-difference time-domain (FDTD) simulations including the sphere and the waveguide test the exact force; multiple scattering between them lowers the depth by 4\,\% (Supplement 1, Sec.~S6). Across the waveguide, the lateral mode profiles close the trap [Fig.~\ref{fig:trap}(b)]. Escaping sideways costs 1.5 times $\Delta U$, so $\Delta U$ is also the depth of the three-dimensional trap (Supplement 1, Sec.~S7).

For a point dipole, the depth relative to the magnitude of the potential at the maximum, $U^\ast=\tfrac14\mathrm{Re}(\alpha)I^\ast$, is $\Delta U/U^\ast=(I^\ast-I_\mathrm{s})/I^\ast$, where $I_\mathrm{s}$ is $|E|^2$ at the saddle that sets $\Delta U$. In the water above the waveguide $\beta_i^2=\kw^2+\gamma_i^2$, so the two decay constants fix the shape of the two-mode field, and the mode amplitudes only scale the field and move it up or down. When the maximum lies well above the surface, $\Delta U/U^\ast$ therefore depends on the decay constants alone (color in Fig.~\ref{fig:map}). It approaches unity when both modes are weakly confined, and along the line of silicon nitride waveguides in Fig.~\ref{fig:map} it exceeds 0.6 near $\gamma_1/\kw\approx0.07$.

The substrate limits how weakly the less confined mode can be confined. On silica ($n_\mathrm{s}=1.45$) both modes need effective indices above 1.45, so $\gamma_1/\kw>0.44$ (dashed line in Fig.~\ref{fig:map}), and TE$_{01}$ is guided only in waveguides thicker than 436~nm. The deepest trap, with the sphere 190~nm above the surface, forms on a 450-nm-thick waveguide [Fig.~\ref{fig:design}(a)], whose modes have $\gamma_0/\kw=0.97$ and $\gamma_1/\kw=0.45$ and meet Eq.~(\ref{eq:criterion}), but $\Delta U/U^\ast$ is only 0.22: the particle escapes along the waveguide, over a saddle, into the neighboring intensity maximum on the surface, at four fifths of the peak intensity.

Without the substrate, the limit on $\gamma_1$ disappears: in a silicon nitride waveguide suspended in water, the cutoff of TE$_{01}$ is set by the water on both sides, and the waveguide can be thinned to 355~nm. A 370-nm-thick waveguide has decay lengths of 139~nm and 1.8~$\mu$m and $\gamma_0\gamma_1/\kw^2=0.064$ [Fig.~\ref{fig:schematic}, inset], and it lifts a maximum 400~nm above its surface [Fig.~\ref{fig:trap}(a)]. The particle now escapes straight down, with $\Delta U/U^\ast=0.68$ for a point dipole at the power ratio of Fig.~\ref{fig:trap}. For the silicon sphere, the trapping potential depth is 0.19~$k_\mathrm{B}T$/mW [Fig.~\ref{fig:design}(a)], or 0.18~$k_\mathrm{B}T$/mW with the effective widths of the modes of the 10-$\mu$m-wide waveguide (5.06 and 5.64~$\mu$m), so about 57~mW give 10~$k_\mathrm{B}T$, enough for stable trapping \cite{Ashkin1986}. Per milliwatt, the stiffness is 0.00025, 0.009 and 0.097~pN/$\mu$m across the waveguide, along the height and along the waveguide; at 57~mW, the position has standard deviations of 0.67, 0.11 and 0.03~$\mu$m about a mean height of 0.44~$\mu$m (Supplement 1, Sec.~S7). The exact depth is 12\,\% below that of the point dipole, mainly because the magnetic dipole of the sphere is pushed away from the lifted maximum of $E_x$, where the magnetic field vanishes. A lifted trap persists for silicon spheres up to about 130~nm in radius and for silica or polystyrene spheres up to 200~nm (Supplement 1, Sec.~S6). At 20~mW the sphere warms by less than 0.1~K, and the water near the waveguide by at most about 0.4~K (Supplement 1, Sec.~S8). Changing the ratio $s=P_0/P_1$ of the TE$_{00}$ to the TE$_{01}$ power shifts the field above the waveguide along $y$ and rescales it: the maximum rises as $\ln s/[2(\gamma_0-\gamma_1)]$, by 173~nm per decade [Fig.~\ref{fig:design}(b)], 1.5 times the standard deviation of the height at 57~mW, and the trap is deepest near $s=0.7$, weakened at larger $s$ by the loss of power in TE$_{01}$ and at smaller $s$ by escape to the surface. The counter-propagating waves need not be balanced: an imbalance of 1.8~dB between the two TE$_{01}$ waves lowers the depth by 1.5\,\%, and one of 3~dB by 10\,\% (Supplement 1, Sec.~S1). Both modes can be fed from one laser through partial TE$_{00}$-to-TE$_{01}$ converters at the two ends; the converters and the length between them then fix the temporal phase between the standing waves, while the phase between the two inputs only moves the trap along the waveguide.
 Because TE$_{01}$ is odd across the thickness, a second, shallower trap ($\Delta U/U^\ast=0.53$) forms at the same height below the waveguide, shifted along it by about half a beat length, $\pi/(\beta_0-\beta_1)=1.13~\mu$m [Fig.~\ref{fig:trap}(a)].
For both waveguides, the trapping potential depth peaks within 15~nm above the cutoff thickness of TE$_{01}$ [Fig.~\ref{fig:design}(a)], because closer to cutoff the mode spreads into the water and its intensity per unit power falls; for the suspended waveguide the peak is 20 times higher than the 0.009~$k_\mathrm{B}T$/mW on silica, and the depth stays above half of it from 358 to 390~nm.

Other substrates and waveguide cores (Supplement 1, Sec.~S5, Table~S1) change little: a substrate of lower index than water deepens the trap by 25\,\% (Teflon AF \cite{Yang2008}) or 29\,\% (sealed air gap), and waveguide cores of lower index hold the particle up to 0.8~$\mu$m above the surface in shallower traps.

With three or more modes, Eq.~(\ref{eq:necessary}) must still hold, and a lifted maximum obeys $E_x\le(1-\gamma_a/\gamma_b)E_+$, with $E_+$ the sum of the positive terms, $\gamma_a$ the smallest decay constant among them and $\gamma_b$ the largest among the canceling terms (Supplement 1, Sec.~S3). With only strongly confined modes, the maximum is weak and needs finely tuned amplitudes, as in Ref.~\cite{Ginis2020}; one weakly confined mode makes it strong (Supplement 1, Fig.~S3).

These results require only the Helmholtz equation for a single-frequency field, a particle drawn to intensity maxima and standing waves that oscillate in phase; they hold for any waveguide material. Traps outside these conditions are not restricted by the budget: particles of negative polarizability sit at intensity minima, which Eq.~(\ref{eq:budget}) does not limit \cite{Ginis2020,Christandl2004,Sague2008}, and two-color traps combine attractive and repulsive potentials from fields of different frequency \cite{Ovchinnikov1991,Vetsch2010}. TM modes, with two field components and transverse spin \cite{Bliokh2014}, need a separate analysis.

In summary, the Helmholtz equation imposes a wavenumber budget on intensity maxima above a waveguide. Two modes, each forming a standing wave, lift a trap off the surface exactly when the product of their decay constants is below the squared wavenumber of the surrounding medium. A deep levitated near-field trap needs a wide, thin waveguide that carries two modes of different order across its thickness, suspended in the surrounding medium or placed on a substrate of lower index, and operated just above the cutoff of the less confined mode. The trap can be tested by imaging a silicon particle above a suspended silicon nitride waveguide: the particle should float only while both modes are launched, at a height that rises by 173~nm per decade of the power ratio. By holding a particle at a chosen height without contact, such traps could measure near-surface forces and the gap dependence of hydrodynamic drag.

\begin{backmatter}
\bmsection{Funding} National Natural Science Foundation of China (12674397); Fundamental Research Funds for the Central Universities (2042026fg0053).
\bmsection{Disclosures} The author declares no conflicts of interest.
\bmsection{Data availability} Data underlying the results presented in this Letter are 
not publicly available at this time but may be obtained from the author upon reasonable request.
\bmsection{Supplemental document} See Supplement 1 for supporting content.
\end{backmatter}

\bibliography{refs}
\end{document}